\documentclass[sigconf]{acmart}

\AtBeginDocument{%
  }

\setcopyright{acmlicensed}
\copyrightyear{2018}
\acmYear{2018}
\acmDOI{XXXXXXX.XXXXXXX}
\acmConference[Conference acronym 'XX]{Make sure to enter the correct
  conference title from your rights confirmation email}{June 03--05,
  2018}{Woodstock, NY}
\acmISBN{978-1-4503-XXXX-X/2018/06}

\begin{document}

\title{Lessons from Biophilic Design: Rethinking Affective Interaction Design in Built Environments}

\author{Shruti Rao}
\email{s.rao@uva.nl}
\affiliation{
    \institution{University of Amsterdam}
    \city{Amsterdam}
    \country{The Netherlands}}

\author{Judith Good}
\email{j.a.good@uva.nl}
\affiliation{%
    \institution{University of Amsterdam}
    \city{Amsterdam}
    \country{The Netherlands}}

\author{Hamed Alavi}
\email{h.alavi@uva.nl}
\affiliation{%
    \institution{University of Amsterdam}
    \city{Amsterdam}
    \country{The Netherlands}}

\renewcommand{\shortauthors}{Rao et al.}

\begin{abstract}
The perspectives of affective interaction in built environments are largely overlooked and instead dominated by affective computing approaches that view emotions as ```static'', computable states to be detected and regulated. To address this limitation, we interviewed architects to explore how biophilic design---our deep-rooted emotional connection with nature---could shape affective interaction design in smart buildings. Our findings reveal that natural environments facilitate self-directed emotional experiences through spatial diversity, embodied friction, and porous sensory exchanges. Based on this, we introduce three design principles for discussion at the Affective Interaction workshop: (1) Diversity of Spatial Experiences, (2) Self-Reflection Through Complexity \& Friction, and (3) Permeability \& Sensory Exchange with the Outside World, while also examining the challenges of integrating these perspectives into built environments.
\end{abstract}



\begin{teaserfigure}
  \includegraphics[width=\textwidth]{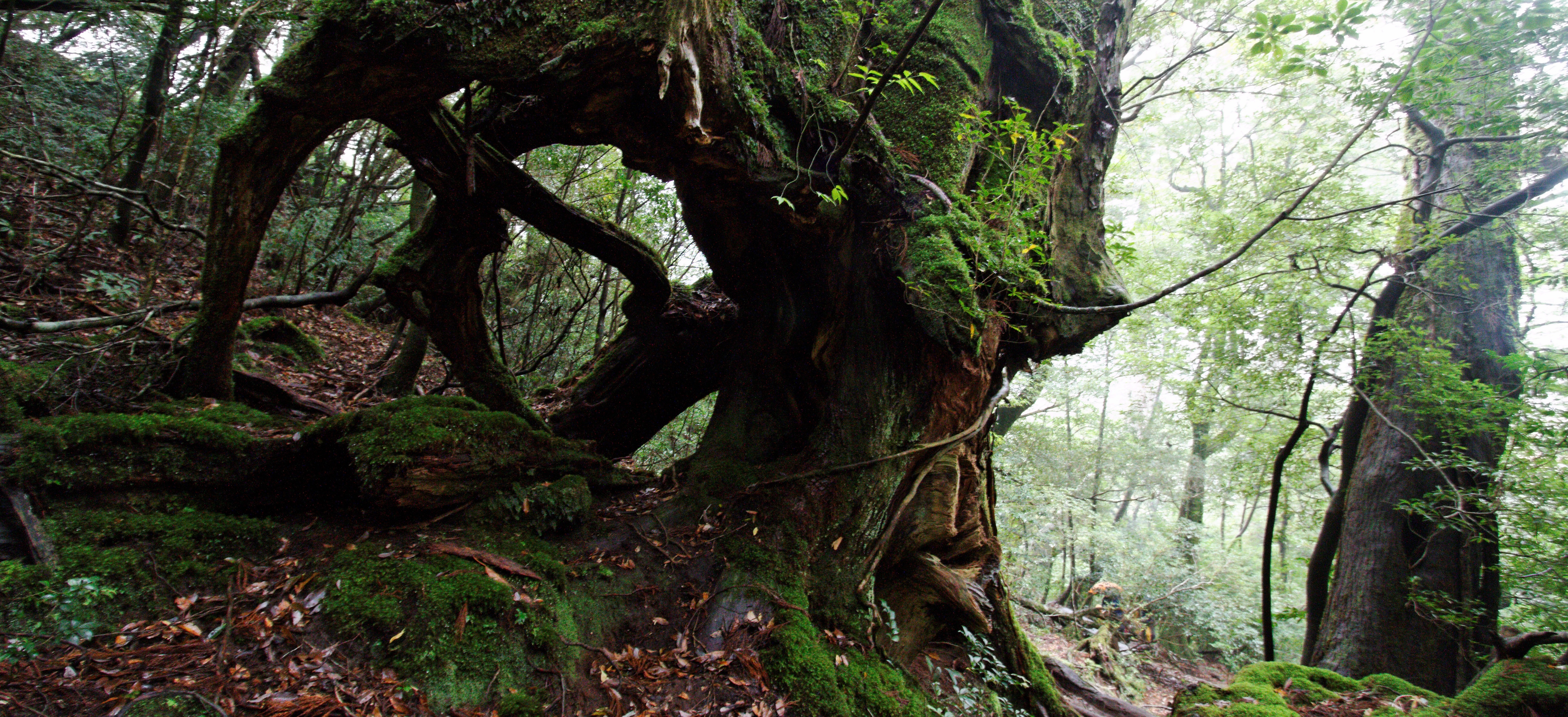}
  \caption{Nature supports emotional exploration through embodied interactions with sensory and spatial conditions. Here, the differing pathways offer opportunities for affective attunement through movement---whether through the more enclosed, shaded tree gaps or along the open forest path. Image Credit: \href{https://commons.wikimedia.org/wiki/File:Shiratani_Unsuikyo_(4196806304).jpg}, Shiratani Unsuikyo, cropped, licensed under \href{https://creativecommons.org/licenses/by-sa/2.0/legalcode}{CC BY-SA 2.0}.}
  \Description{A mossy forest in Japan}
  \label{fig:teaser}
\end{teaserfigure}

\received{20 February 2007}
\received[revised]{12 March 2009}
\received[accepted]{5 June 2009}

\maketitle

\section{Introduction}
Built environments are increasingly designed to support a range of human experiences such as comfort~\cite{alavi2017comfort} and socialisation~\cite{nguyen2024adaptive}, yet their approach to emotions remains largely reductive~\cite{ghandi2021embodied, yadav2018capturing}. These environments often rely on affective computing approaches, which frame emotions as computable states to be detected (e.g. through valence-arousal models~\cite{gao2020n, gao2022understanding} or machine-readable cues~\cite{caon2014affective}) and subsequently regulated. For instance, smart lighting systems may adjust colour temperature to induce relaxation upon detecting stress in occupants~\cite{laushkina2020implementation, cupkova2019intelligent}. 

In contrast, affective interaction~\cite{hook2008knowing} emphasises emotions as emergent from ongoing, situated interactions with technology. While this research has explored how technology supports meaning-making through emotion, it has primarily focused on digital and tangible interfaces (e.g. \cite{breazeal2001affective, zhou2024tangible}), leaving built environments largely overlooked despite their fundamental role in emotional experiences.

To address this gap, this paper explores how insights from biophilic design---an architectural approach rooted in our innate emotional connection with nature---can inform affective interaction in built environments. This paper builds on our primary research, in which we interviewed 13 architects to investigate how biophilic design can shape human interactions in future smart buildings~\cite{wilson1986biophilia, kellert1995biophilia, kellert2018nature}. While we did not originally frame this work through the lens of affective interaction, re-analysing our data revealed that biophilic design offers valuable insights for affective interaction in built environments. 

Our findings suggest that natural environments provide sensory and spatial experiences that guide us in attuning to, discovering, and making sense of our emotions~\cite{kellert2018nature, joye2018nature}. This leads us to our position:\emph{
smart buildings should not be designed to elicit specific emotional outcomes or regulate emotional states. Instead, they should provide conditions that empower individuals to sense, reflect on, and navigate their own emotions in ways that feel natural---through ongoing, moment-to-moment interactions with the built environment.} One architect captured this perspective: \textit{```(in nature) the body intuitively finds balance in experiences---for instance, openness after enclosure, groundedness after exposure----doesn't matter what that is, as long as we (in buildings) can provide support for the body to get that balance...''}.

\section{Designing for Situated and Emergent Emotional Experiences in Built Environments}

Based on the interview data ($N=13$), we identified three design messages---conditions (that we mentioned in our position)---that can enable affective interaction in built environments, drawing inspiration from how emotions emerge in biophilic settings. Unlike smart spaces that optimise for predefined emotional states (e.g. promoting focus through automated lighting), natural environments allow emotions to unfold through situated interactions. Translating this into built environments requires shifting from controlling emotional states to enabling users to intuitively engage with their surroundings based on evolving needs. This shift positions the body as a sensor, an active interface that continuously interprets and responds to environmental cues.

\paragraph{Diversity of Spatial Experiences:} 
Natural environments provide diverse spatial conditions that individuals navigate based on their evolving needs. In built environments, this means allowing users to self-select and transition between different spatial experiences that align with their momentary states. As one architect reflected: \emph{```You need a certain connection (in reference to emotion)---not everybody, and not always, and not in the same amount. On the other hand, you need a certain groundedness---like you can step out of the connection and be by yourself in a different state of mind.''} As an example, building technologies can support spatial interactions that allow users to modulate their sense of connection and solitude. Somatic thresholds---interactive transitions with soft, responsive barriers---could enable users to physically shift between openness and enclosure based on how they engage with their surroundings. If a person pauses near the threshold, the opening could subtly contract, creating a more enclosed, introspective space. If they step forward, the partition could expand, inviting more openness and connection. Engaging with these spatial shifts creates gentle environmental feedback loops that prompt users to notice and reflect on their behaviour---whether they are seeking solitude, openness, or an in-between state. By engaging with these transitions, individuals can explore how their presence, movement, and spatial positioning influence their emotions.

\paragraph{Self-Reflection Through Complexity \& Friction:} Natural environments can be challenging, often presenting moments of friction that require individuals to accept and adapt (e.g. a sudden storm can transform an initially pleasant hiking experience). Dealing with such frictions can compel people to attune to their bodies and emotions, catalysing personal reflection and emotional awareness. As one expert noted, \textit{```Friction fosters reflection. When we’re uncomfortable, we’re more likely to pay attention to our surroundings, and find meaning in our experience...(to accept) our circumstances.''} In contrast, many smart buildings prioritise supporting  uniform emotions such as calm or productivity by smoothing out variations in climate, sound, and spatial experience. While this optimises comfort, it can also diminish opportunities for self-learning and self-regulation. For example, technology in workspaces can explore the use of  lighting glares in social spaces that momentarily interrupt occupants from being too involved with their screens and instead engage with people or the environment around them. Rather than prescribing emotional responses, these designed frictions create ```affective touchpoints'', prompting users to sense, interpret, and navigate their own evolving states within the environment.

\paragraph{Permeability \& Sensory Exchange with the Outside World:} Biophilic experiences are inherently porous, allowing for continuous exchange between interior and exterior conditions---light, air, sound, and movement. Many buildings, however, are designed as sealed environments, limiting exposure to natural variation and seasonal rhythms that provide implicit temporal cues for emotional regulation. As one expert noted, \textit{```When we open windows, we don’t just let in air---we let in time, season, meaning, and place.''} As an example, smart offices could incorporate a sliding panel that users can physically adjust to modulate their exposure to outdoor sounds throughout the day. Variations in ambient sounds such as distant evening traffic can create an intuitive sense of time, helping users remain anchored in the present moment. By adjusting their connection to these external cues, users can develop a felt awareness of temporal shifts, realising when it’s time to pause, refocus, or transition between tasks. Rather than passively experiencing a predefined atmosphere, this interactive interface encourages an active, embodied negotiation of emotional states, allowing individuals to shape their emotional experience through their relationship with sensory conditions.

\section{Challenges and Conclusion}
Altogether, the discussed design principles shape an approach to ```affective built environments''~\cite{rao2023towards} that is grounded in bodily interaction, supports self-directed emotional navigation, and emerges through spatial and sensory engagement with the environment. Integrating affective interactions into built environments presents several challenges---both in applying the discussed design principles and in addressing broader questions about affective interaction. These are key challenges we aim to explore at the Affective Interaction workshop: (1) \textbf{Designing for Open-Ended Affective Experiences} — How can we design environments that offer users the freedom to navigate their emotional experiences intuitively while capturing subtle, embodied responses and decisions made? (2) \textbf{Embedding Embodied Intelligence in Technology} — How can technology surface, support, and amplify bodily intuition without abstracting it into reductive emotional models? (3) \textbf{Balancing Adaptation and Non-Intrusiveness} — How can environments interact with individual (and collective) emotional states in non-intrusive ways, preserving user agency and emergent emotional meaning-making?

Nature supports an intuitive process of emotional exploration, yet built environments fail to provide the same affordances. By applying biophilic principles, we emphasise the need for affective interaction in built environments and explore how affective meaning-making can be embedded in the spaces we inhabit. Additionally, we put forth considerations on how technology can support this process without overriding the fluid, emergent nature of emotional experiences.

\bibliographystyle{ACM-Reference-Format}
\bibliography{sample-base}


\begin{thebibliography}{17}


\ifx \showCODEN    \undefined \def \showCODEN     #1{\unskip}     \fi
\ifx \showISBNx    \undefined \def \showISBNx     #1{\unskip}     \fi
\ifx \showISBNxiii \undefined \def \showISBNxiii  #1{\unskip}     \fi
\ifx \showISSN     \undefined \def \showISSN      #1{\unskip}     \fi
\ifx \showLCCN     \undefined \def \showLCCN      #1{\unskip}     \fi
\ifx \shownote     \undefined \def \shownote      #1{#1}          \fi
\ifx \showarticletitle \undefined \def \showarticletitle #1{#1}   \fi
\ifx \showURL      \undefined \def \showURL       {\relax}        \fi
\providecommand\bibfield[2]{#2}
\providecommand\bibinfo[2]{#2}
\providecommand\natexlab[1]{#1}
\providecommand\showeprint[2][]{arXiv:#2}

\bibitem[Alavi et~al\mbox{.}(2017)]%
        {alavi2017comfort}
\bibfield{author}{\bibinfo{person}{Hamed~S Alavi}, \bibinfo{person}{Himanshu Verma}, \bibinfo{person}{Michael Papinutto}, {and} \bibinfo{person}{Denis Lalanne}.} \bibinfo{year}{2017}\natexlab{}.
\newblock \showarticletitle{Comfort: A coordinate of user experience in interactive built environments}. In \bibinfo{booktitle}{\emph{IFIP conference on human-computer interaction}}. Springer, \bibinfo{publisher}{Springer}, \bibinfo{address}{Mumbai, India}, \bibinfo{pages}{247--257}.
\newblock


\bibitem[Breazeal(2001)]%
        {breazeal2001affective}
\bibfield{author}{\bibinfo{person}{Cynthia Breazeal}.} \bibinfo{year}{2001}\natexlab{}.
\newblock \showarticletitle{Affective interaction between humans and robots}. In \bibinfo{booktitle}{\emph{European conference on artificial life}}. Springer, \bibinfo{pages}{582--591}.
\newblock


\bibitem[Caon et~al\mbox{.}(2014)]%
        {caon2014affective}
\bibfield{author}{\bibinfo{person}{Maurizio Caon}, \bibinfo{person}{Leonardo Angelini}, \bibinfo{person}{Omar Abou~Khaled}, \bibinfo{person}{Denis Lalanne}, \bibinfo{person}{Yong Yue}, {and} \bibinfo{person}{Elena Mugellini}.} \bibinfo{year}{2014}\natexlab{}.
\newblock \showarticletitle{Affective interaction in smart environments}.
\newblock \bibinfo{journal}{\emph{Procedia Computer Science}}  \bibinfo{volume}{32} (\bibinfo{year}{2014}), \bibinfo{pages}{1016--1021}.
\newblock


\bibitem[Cupkova et~al\mbox{.}(2019)]%
        {cupkova2019intelligent}
\bibfield{author}{\bibinfo{person}{Dominika Cupkova}, \bibinfo{person}{Erik Kajati}, \bibinfo{person}{Jozef Mocnej}, \bibinfo{person}{Peter Papcun}, \bibinfo{person}{Jiri Koziorek}, {and} \bibinfo{person}{Iveta Zolotova}.} \bibinfo{year}{2019}\natexlab{}.
\newblock \showarticletitle{Intelligent human-centric lighting for mental wellbeing improvement}.
\newblock \bibinfo{journal}{\emph{International Journal of Distributed Sensor Networks}} \bibinfo{volume}{15}, \bibinfo{number}{9} (\bibinfo{year}{2019}), \bibinfo{pages}{1550147719875878}.
\newblock


\bibitem[Gao et~al\mbox{.}(2022)]%
        {gao2022understanding}
\bibfield{author}{\bibinfo{person}{Nan Gao}, \bibinfo{person}{Max Marschall}, \bibinfo{person}{Jane Burry}, \bibinfo{person}{Simon Watkins}, {and} \bibinfo{person}{Flora~D Salim}.} \bibinfo{year}{2022}\natexlab{}.
\newblock \showarticletitle{Understanding occupants’ behaviour, engagement, emotion, and comfort indoors with heterogeneous sensors and wearables}.
\newblock \bibinfo{journal}{\emph{Scientific Data}} \bibinfo{volume}{9}, \bibinfo{number}{1} (\bibinfo{year}{2022}), \bibinfo{pages}{1--16}.
\newblock


\bibitem[Gao et~al\mbox{.}(2020)]%
        {gao2020n}
\bibfield{author}{\bibinfo{person}{Nan Gao}, \bibinfo{person}{Wei Shao}, \bibinfo{person}{Mohammad~Saiedur Rahaman}, {and} \bibinfo{person}{Flora~D Salim}.} \bibinfo{year}{2020}\natexlab{}.
\newblock \showarticletitle{n-gage: Predicting in-class emotional, behavioural and cognitive engagement in the wild}.
\newblock \bibinfo{journal}{\emph{Proceedings of the ACM on Interactive, Mobile, Wearable and Ubiquitous Technologies}} \bibinfo{volume}{4}, \bibinfo{number}{3} (\bibinfo{year}{2020}), \bibinfo{pages}{1--26}.
\newblock


\bibitem[Ghandi et~al\mbox{.}(2021)]%
        {ghandi2021embodied}
\bibfield{author}{\bibinfo{person}{Mona Ghandi}, \bibinfo{person}{Marcus Blaisdell}, {and} \bibinfo{person}{Mohamed Ismail}.} \bibinfo{year}{2021}\natexlab{}.
\newblock \showarticletitle{Embodied empathy: Using affective computing to incarnate human emotion and cognition in architecture}.
\newblock \bibinfo{journal}{\emph{International Journal of Architectural Computing}} \bibinfo{volume}{19}, \bibinfo{number}{4} (\bibinfo{year}{2021}), \bibinfo{pages}{532--552}.
\newblock


\bibitem[Hook(2008)]%
        {hook2008knowing}
\bibfield{author}{\bibinfo{person}{Kristina Hook}.} \bibinfo{year}{2008}\natexlab{}.
\newblock \showarticletitle{Knowing, communication and experiencing through body and emotion}.
\newblock \bibinfo{journal}{\emph{IEEE Transactions on Learning technologies}} \bibinfo{volume}{1}, \bibinfo{number}{4} (\bibinfo{year}{2008}), \bibinfo{pages}{248--259}.
\newblock


\bibitem[Joye and Dewitte(2018)]%
        {joye2018nature}
\bibfield{author}{\bibinfo{person}{Yannick Joye} {and} \bibinfo{person}{Siegfried Dewitte}.} \bibinfo{year}{2018}\natexlab{}.
\newblock \showarticletitle{Nature's broken path to restoration. A critical look at Attention Restoration Theory}.
\newblock \bibinfo{journal}{\emph{Journal of environmental psychology}}  \bibinfo{volume}{59} (\bibinfo{year}{2018}), \bibinfo{pages}{1--8}.
\newblock


\bibitem[Kellert(2018)]%
        {kellert2018nature}
\bibfield{author}{\bibinfo{person}{Stephen~R Kellert}.} \bibinfo{year}{2018}\natexlab{}.
\newblock \bibinfo{booktitle}{\emph{Nature by design: The practice of biophilic design}}.
\newblock \bibinfo{publisher}{yale university press}.
\newblock


\bibitem[Kellert and Wilson(1995)]%
        {kellert1995biophilia}
\bibfield{author}{\bibinfo{person}{Stephen~R Kellert} {and} \bibinfo{person}{Edward~O Wilson}.} \bibinfo{year}{1995}\natexlab{}.
\newblock \showarticletitle{The biophilia hypothesis}.
\newblock  (\bibinfo{year}{1995}).
\newblock


\bibitem[Laushkina et~al\mbox{.}(2020)]%
        {laushkina2020implementation}
\bibfield{author}{\bibinfo{person}{AA Laushkina}, \bibinfo{person}{SV Roslyakova}, {and} \bibinfo{person}{AV Smirnov}.} \bibinfo{year}{2020}\natexlab{}.
\newblock \showarticletitle{Implementation of adaptive lighting systems to reduce stressful situations in multi-user spaces}.
\newblock \bibinfo{journal}{\emph{Information technologies}} \bibinfo{volume}{5}, \bibinfo{number}{4} (\bibinfo{year}{2020}), \bibinfo{pages}{62--69}.
\newblock


\bibitem[Nguyen and Vande~Moere(2024)]%
        {nguyen2024adaptive}
\bibfield{author}{\bibinfo{person}{Binh Vinh~Duc Nguyen} {and} \bibinfo{person}{Andrew Vande~Moere}.} \bibinfo{year}{2024}\natexlab{}.
\newblock \showarticletitle{The Adaptive Architectural Layout: How the Control of a Semi-Autonomous Mobile Robotic Partition was Shared to Mediate the Environmental Demands and Resources of an Open-Plan Office}. In \bibinfo{booktitle}{\emph{Proceedings of the CHI Conference on Human Factors in Computing Systems}}. \bibinfo{pages}{1--20}.
\newblock


\bibitem[Rao et~al\mbox{.}(2023)]%
        {rao2023towards}
\bibfield{author}{\bibinfo{person}{Shruti Rao}, \bibinfo{person}{Hamed Alavi}, {and} \bibinfo{person}{Judith Good}.} \bibinfo{year}{2023}\natexlab{}.
\newblock \showarticletitle{Towards Empathic Buildings: Exploring How Smart Buildings May Be Designed to Address Occupants’ Subjective Needs}. In \bibinfo{booktitle}{\emph{Proceedings of the 2nd Empathy-Centric Design Workshop}}. \bibinfo{pages}{1--4}.
\newblock


\bibitem[Wilson(1986)]%
        {wilson1986biophilia}
\bibfield{author}{\bibinfo{person}{Edward~O Wilson}.} \bibinfo{year}{1986}\natexlab{}.
\newblock \bibinfo{booktitle}{\emph{Biophilia}}.
\newblock \bibinfo{publisher}{Harvard university press}.
\newblock


\bibitem[Yadav et~al\mbox{.}(2018)]%
        {yadav2018capturing}
\bibfield{author}{\bibinfo{person}{Megha Yadav}, \bibinfo{person}{Theodora Chaspari}, \bibinfo{person}{Jinwoo Kim}, {and} \bibinfo{person}{Changbum~R Ahn}.} \bibinfo{year}{2018}\natexlab{}.
\newblock \showarticletitle{Capturing and quantifying emotional distress in the built environment}. In \bibinfo{booktitle}{\emph{Proceedings of the Workshop on Human-Habitat for Health (H3): Human-Habitat Multimodal Interaction for Promoting Health and Well-Being in the Internet of Things Era}}. \bibinfo{pages}{1--8}.
\newblock


\bibitem[Zhou et~al\mbox{.}(2024)]%
        {zhou2024tangible}
\bibfield{author}{\bibinfo{person}{Nianmei Zhou}, \bibinfo{person}{Steven Devleminck}, {and} \bibinfo{person}{Luc Geurts}.} \bibinfo{year}{2024}\natexlab{}.
\newblock \showarticletitle{Tangible affect: A literature review of tangible interactive systems addressing human core affect, emotions and moods}. In \bibinfo{booktitle}{\emph{Proceedings of the 2024 ACM Designing Interactive Systems Conference}}. \bibinfo{pages}{424--440}.
\newblock


\end{thebibliography}

\end{document}